\documentclass[%
	final,
   journal,
    comsoc,
	letterpaper,
	oneside,
	twocolumn,
	nofonttune,%
]{IEEEtran}%
\PassOptionsToPackage{usenames,dvipsnames,svgnames,x11names,table,prologue}{xcolor}%
\PassOptionsToPackage{hyphens}{url}%
\PassOptionsToPackage{nomessages}{fp}%
\ifCLASSINFOpdf
  \usepackage[pdftex]{graphicx}
\else
  \usepackage[dvips]{graphicx}
\fi
\ifCLASSOPTIONcompsoc
   \usepackage[caption=false,font=normalsize,labelfont=sf,textfont=sf]{subfig}
\else
   \usepackage[caption=false,font=footnotesize]{subfig}
\fi

\usepackage{stfloats}
\usepackage[english]{babel}%
\usepackage[utf8]{inputenc}%
\usepackage[babel,style=english]{csquotes}%
\usepackage{hyphsubst}%
\usepackage[%
	activate={true,%
	nocompatibility},%
	final,%
	tracking=true,%
	kerning=true,%
	spacing=true,%
	factor=1100,%
	stretch=10,%
	shrink=10%
]{microtype}%
\usepackage{setspace}%
\usepackage{xcolor}%
\definecolor{todonotecol}{RGB}{250,0,0}%
\usepackage{xparse}
\usepackage[%
	colorlinks=false,%
	urlcolor=black,%
	linkcolor=black,%
	citecolor=black,%
	filecolor=black,%
	breaklinks,%
	]{hyperref}%
\usepackage{url}%
\usepackage[%
	acronym,%
	nopostdot,%
	seeautonumberlist,%
	shortcuts,%
	section=chapter,%
	toc,%
]{glossaries}%
\glsaddkey*{url}
{}
{\glsentryurl}
{\Glsentryurl}
{\glsurl}
{\Glsurl}
{\GLSurl}
\glsaddkey*{longpltwo}
{}
{\glsentrylongpltwo}
{\Glsentrylongpltwo}
{\glslongpltwo}
{\Glslongpltwo}
{\GLSlongpltwo}
\newglossaryentry{}{%
	name={},%
	plural={},%
	description={},%
}%
\newacronym{dmz}{DMZ}{demilitarized zone}
\newacronym{it}{IT}{information technology}
\newacronym{ot}{OT}{operational technology}
\newacronym{opcua}{OPC UA}{Open Platform Communications Unified Architecture}
\newacronym{tsn}{TSN}{time-sensitive networking}
\newacronym{scada}{SCADA}{supervisory control and data acquisition}
\newacronym{vpc}{VPC}{Virtual Process Controller}
\newacronym{plc}{PLC}{programmable logic controller}
\newacronym{udp}{UDP}{User Datagram Protocol}
\newacronym{pubsub}{PubSub}{Publish/Subscribe}
\newacronym{vm}{VM}{virtual machine}
\newacronym{pc}{PC}{personal computer}
\newacronym{ptp}{PTP}{Precision Time Protocol}
\newacronym{ntp}{NTP}{Network Time Protocol}
\newacronym{splc}{Soft-PLC}{Software-PLC}
\newacronym{rte}{RTE}{Real-Time Ethernet}
\newacronym{osi}{OSI}{Open Systems Interconnection}
\newacronym{mac}{MAC}{Media Access Control}
\newacronym{e2e}{E2E}{end-to-end}
\newacronym{tcp}{TCP}{Transmission Control Protocol}
\newacronym{hmi}{HMI}{Human-Machine Interface}
\newacronym{kpi}{KPI}{Key Performance Indicator}
\newacronym{mes}{MES}{Manufacturing Execution System}
\newacronym{rami4.0}{RAMI 4.0}{Reference Architectural Model Industrie 4.0}
\newacronym{tacnet4.0}{TACNET 4.0}{Tactile Internet 4.0}
\newacronym{3gpp}{3GPP}{3rd Generation Partnership Project}
\newacronym{5gppp}{5GPPP}{5G Infrastructure Public Private Partnership}
\newacronym{itu}{ITU}{International Telecommunication Union}
\newacronym{ngnm}{NGNM}{Next Generation Mobile Networks}
\newacronym{agv}{AGV}{automated guided vehicle}
\newacronym{v2v}{V2V}{vehicle-to-vehicle}
\newacronym{v2x}{V2X}{vehicle-to-infrastructure}
\newacronym{d2x}{D2X}{device-to-x}
\newacronym{qos}{QoS}{quality of service}
\newacronym{noa}{NOA}{NAMUR Open Architecture}
\newacronym{dcs}{DCS}{distributed control system}
\newacronym{m2m}{M2M}{machine-to-machine}
\newacronym{sdn}{SDN}{software-defined networking}
\newacronym{erp}{ERP}{enterprise resource planning}
\newacronym{ip}{IP}{Internet Protocol}
\newacronym{lte}{LTE}{Long Term Evolution}
\newacronym{5g}{5G}{5th generation wireless communication system}
\newacronym{harq}{HARQ}{Hybrid Automatic Repeat Request}
\newacronym{rat}{RAT}{radio access technology}
\newacronym{mc}{MC}{Multi-Connectivity}
\newacronym{fbmc}{FBMC}{Filter Bank Multicarrier}
\newacronym{gfdm}{GFDM}{Generalized Frequency Division Multiplexing}
\newacronym{sla}{SLA}{service-level agreement}
\newacronym{5qi}{5QI}{5G QoS Indicator}
\newacronym{bmbf}{BMBF}{German Federal Ministry of Education and Research}
\newacronym{nrt}{NRT}{non real-time}
\newacronym{irt}{IRT}{isochronous real-time}
\newacronym{rt}{RT}{real-time}
\newacronym{ar}{AR}{augmented reality}
\newacronym{wlan}{WLAN}{wireless local area network}
\newacronym{lan}{LAN}{local area network}
\newacronym{cpu}{CPU}{central processing unit}
\newacronym{ram}{RAM}{random-access memory}
\newacronym{ue}{UE}{user equipment}
\newacronym{ran}{RAN}{radio access network}
\newacronym{bs}{BS}{base station}
\newacronym{cn}{CN}{core network}
\newacronym{epc}{EPC}{evolved packet core}
\newacronym{mec}{MEC}{mobile edge cloud}
\newacronym{rtt}{RTT}{round-trip time}
\newacronym{vpn}{VPN}{virtual private network}
\newacronym{vpf}{VPF}{Virtual Process Control Function}
\newacronym{vdi}{VDI}{Association of German Engineers}
\newacronym{ipc}{IPC}{industrial pc}
\newacronym{rtos}{RTOS}{real-time operating system}
\newacronym{vpcmo}{VPC M\&O}{VPC Management \& Orchestration}
\newacronym{i/o}{I/O}{input /output}
\newacronym{fbd}{FBD}{Function Block Diagram}
\newacronym{ldd}{LDD}{Ladder Logic Diagram}
\newacronym{sfc}{SFC}{Sequential Function Chart}
\newacronym{il}{IL}{Instruction List}
\newacronym{st}{ST}{Structured Text}
\newacronym{os}{OS}{operating system}
\newacronym{ir}{IR}{Inactive Resource}
\newacronym{msc}{MSC}{message sequence chart}
\newacronym{icps}{ICPS}{industrial cyber-physical system}
\newacronym{cps}{CPS}{cyber-physical systems}
\newacronym{mtd}{MTD}{Moving Target Defense}
\newacronym{iot}{IoT}{Internet of Things}
\newacronym{iiot}{IIoT}{industrial Internet of Things}
\newacronym{iiot2}{IIoT}{Industrial IoT}
\newacronym{k8s}{K8s}{Kubernetes}
\newacronym{rkt}{rkt}{CoreOS Rocket}
\newacronym{dcos}{DC/OS}{Distributed Cloud Operating System}
\newacronym{ml}{ML}{machine learning}
\newacronym{ai}{AI}{artificial intelligence}
\newacronym{ict}{ICT}{industrial communication technology}
\newacronym{kvm}{KVM}{kernel-based virtual machine}
\newacronym{ie}{IE}{Industrial Ethernet}
\newacronym{ias}{IAS}{industrial automation system}
\newacronym{cots}{COTS}{commercial off-the-shelf}
\newacronym{slat}{SLAT}{Second Level Address Translation}
\newacronym{gui}{GUI}{graphical user interface}
\newacronym{l2}{L2}{layer 2}
\newacronym{l3}{L3}{layer 3}
\newacronym{ouc}{OUC}{Open User Communication}
\newacronym{api}{API}{application programming interface }
\newacronym{fb}{FB}{function block}
\newacronym{mqtt}{MQTT}{message queuing telemetry transport }
\newacronym{amqp}{AMQP}{advanced message queuing protocol}
\newacronym{embb}{eMBB}{enhanced mobile broadband}
\newacronym{urllc}{URLLC}{ultra-reliable low-latency communication}
\newacronym{mmtc}{mMTC}{massive machine-type communications}
\newacronym{npn}{NPN}{non-public network}
\newacronym{prb}{PRB}{Physical Ressource Block}
\newacronym{gnb}{gNB}{5G base station}
\newacronym{5gs}{5GS}{5G system}
\newacronym{5gc}{5GC}{5G core network}
\newacronym{oai}{OAI}{OpenAirInterface}
\newacronym{enb}{eNB}{4G base station}
\newacronym{sdr}{SDR}{Software Defined Radio}
\newacronym{usrp}{USRP}{Universal Software Radio Peripheral}
\newacronym{mme}{MME}{Mobility Management Entity}
\newacronym{hss}{HSS}{Home Subscriber Server}
\newacronym{pdn}{PDN}{Packet Data Network}
\newacronym{sgw}{SGW}{Serving Gateway}
\newacronym{gtp}{GTP}{GPRS Tunneling Protocol}
\newacronym{mimo}{MIMO}{Multiple Input Multiple Output}
\newacronym{fpga}{FPGA}{Field Programmable Gate Array}
\newacronym{pss}{PSS}{primary synchronization signal}
\newacronym{sss}{SSS}{secondary synchronization signal}
\newacronym{sfn}{SFN}{system frame number}
\newacronym{mib}{MIB}{master information block}
\newacronym{pbch}{PBCH}{Physical Broadcast Channel}
\newacronym{ism}{ISM}{Industrial, Scientific and Medical}
\newacronym{e-utra}{E-UTRA}{evolved UMTS Terrestrial Radio Access}
\newacronym{bldc}{BLDC}{brushless DC}
\newacronym{dl}{DL}{Downlink}
\newacronym{ds-tt}{DS-TT}{Device-Side TSN Translator}
\newacronym{nw-tt}{NW-TT}{Network-Side TSN Translator}
\newacronym{upf}{UPF}{User Plane Function}
\newacronym{tsi}{TSi}{ingress timestamping}
\newacronym{tse}{TSe}{egress timestamping}
\newacronym{gm}{GM}{grandmaster}
\newacronym{soap}{SOAP}{Simple Object Access Protocol}
\newacronym{http}{HTTP}{Hypertext Transfer Protocol}
\newacronym{pdu}{PDU}{protocol data unit}
\newacronym{flop}{FLOP}{floating point operation}
\glsdisablehyper
\usepackage[%
	backend=biber,%
	style=ieee,%
	isbn=false,%
	hyperref=true,%
	maxbibnames=99,%
	sorting=none,%
	natbib=true,%
	language=english,%
	defernumbers=true,%
	]{biblatex}%
\DeclareFieldFormat{sentencecase}{\csname bbx@colon@search\endcsname#1}

\usepackage{nameref}
\usepackage{soul}
\usepackage{balance}
\usepackage{textcomp}
\usepackage{calc}
\usepackage{xkeyval}
\usepackage{multirow}
\usepackage{tabularx}
\usepackage{tabulary}
\usepackage{makecell}
\usepackage{verbatim}%
\usepackage{pgfplots}
\usepackage{tikz}
\usepackage{textcomp} 
\newcommand{\nl}{\par\noindent} 
\newcommand{\mytilde}{{\raise.17ex\hbox{$\scriptstyle\mathtt{\sim}$}}}

\newlength\textheighttemp%
\newlength\textwidthtemp%
\newlength\textheightstd%
\newlength\textwidthstd%
\newlength\textheightold%
\newlength\textwidthold%
\newlength\tempheight%
\newlength\tempwidth%
\SetProtrusion{encoding={*},family={bch},series={*},size={6,7}}
              {1={ ,750},2={ ,500},3={ ,500},4={ ,500},5={ ,500},
               6={ ,500},7={ ,600},8={ ,500},9={ ,500},0={ ,500}}
\SetExtraKerning[unit=space]
    {encoding={*}, family={bch}, series={*}, size={footnotesize,small,normalsize}}
    {\textendash={400,400}, 
     "28={ ,150}, 
     "29={150, }, 
     \textquotedblleft={ ,150}, 
     \textquotedblright={150, }} 
\SetTracking{encoding={*}, shape=sc}{40}
\makeatletter
\let\blx@rerun@biber\relax
\makeatother

\usepackage{newtxmath}
\usetikzlibrary{external}
\pgfplotsset{
  grid style = {
   line width = 0.1pt
  }
}
\definecolor{blue}{RGB}{0,68,170}%
				\newcommand{\disablewr}[1]{#1}%
				\newcommand{\newcommanddisw}[3]{\newcommand{#1}[1]{\disablewr{\textcolor{#2}{#3}}}}%
\newcommand{\hide}[1]{}

\definecolor{todocol}{named}{red}
\newcommanddisw{\todo}{todocol}{ToDo: #1}%
\definecolor{migucol}{named}{purple}%
\newcommanddisw{\migucom}{migucol}{{@}comment: #1}%
\newcommanddisw{\miguhigh}{migucol}{#1}%
\definecolor{josccol}{named}{brown}%
\newcommanddisw{\josccom}{josccol}{{@}comment: #1}%
\newcommanddisw{\joschigh}{josccol}{#1}%

\begin{document}%
%

\title{
Computation Offloading at Field Level: Motivation and Break-Even Point Calculation 
\thanks{This research was supported by the German Federal Ministry for Economic Affairs and Energy (BMWi) within the project FabOS under grant number 01MK20010C and by the German Federal Ministry of Education and Research (BMBF) within the TACNET~4.0 project under grant number 16KIS0712K. The responsibility for this publication lies with the authors. This is a preprint of a work accepted but not yet published at the IEEE 26th International Conference on Emerging Technologies and Factory Automation (ETFA). Please cite as: M. Gundall, C. Huber, and H.D. Schotten: “Computation Offloading at Field Level: Motivation and Break-Even Point Calculation”. In: 2021 IEEE 26th International Conference on Emerging Technologies and Factory Automation (ETFA), IEEE, 2021.}%
}%
%
%
\author{%
\IEEEauthorblockN{%
    Michael Gundall\IEEEauthorrefmark{1}, %
    Christopher Huber\IEEEauthorrefmark{1}, %
    and Hans D. Schotten\IEEEauthorrefmark{1}\IEEEauthorrefmark{2} %
    \\%
}%
\IEEEauthorblockA{%
    \IEEEauthorrefmark{1}German Research Center for Artificial Intelligence GmbH (DFKI), Kaiserslautern, Germany \\%
    \IEEEauthorrefmark{2}Department of Electrical and Computer Engineering, Technische Universität Kaiserslautern, Kaiserslautern, Germany %
	\\%
    Email: %
        \{michael.gundall, christopher.huber, hans\_dieter.schotten%
        \}@dfki.de
        \\%
}%
}%


%

%
%
%
%
%
%
%
%
\maketitle
%
%
%
%
%
\begin{abstract}%
Smart manufacturing has the objective of creating highly flexible and resource optimized industrial plants. Furthermore, the improvement of product quality is another important target. These requirements implicate more complex control algorithms. Processing these algorithms may exceed the capabilities of resource constrained devices, such as \glspl{plc}. 
In this case, the necessity for computation offloading is given. Due to the fact that industrial plants are currently designed for a life-cycle-time of more than ten years, in a realistic smart manufacturing scenario, these devices have to be considered. Therefore, we investigate the impact of complex algorithms on conventional \glspl{plc} by simulating them with a load generator.

In addition, we propose a realistic factory scenario including benchmarks for both wireline and wireless communication systems. Thus, their \gls{rtt} is measured with and without additional load on the network. With the help of these investigations, break-even points for the application of computation offloading of two typical \glspl{plc} of Siemens S7 series can be calculated. 

\end{abstract}%
\begin{IEEEkeywords}
Computation offloading, PLC, smart manufacturing, Industry 4.0, industrial communication, wireless communications, cloud computing
\end{IEEEkeywords}
%
%
%
%
%
\IEEEpeerreviewmaketitle
%
%
%
%
%
%
%
%
\section{Introduction}%
\label{sec:Introduction}

Today's industrial production lines are controlled by \glspl{plc}. These systems are sophisticated for certain control tasks, but do not provide flexibility and only low computational power. In addition, with the vision of Industry 4.0, many novel use cases have emerged that bring up challenges and requirements that are different from the state of the art \cite{gundall20185g}. Besides mobility requirements imposed by the increasing number of mobile devices, \gls{ml} and \gls{ai} will also have an important role in novel industrial facilities. Complex algorithms need to be solved to apply these methods. Here, the computing power of legacy field devices, such as \glspl{plc}, is not sufficient. 

Following the return on investment and profitability, industrial plants are usually designed for a life-cycle-time of ten years or more \cite{5535166}. Since changes to the plant hardware involve high costs and long holding times, it is not common to upgrade the entire plant to novel technologies and equipment during this time. Therefore, a way to integrate novel technologies as well as a migration path is a must for the so-called brownfield scenarios. To overcome the bottleneck of limited computing power, complex and time-consuming computations can be moved to an edge device. Thus, the limitations of conventional \glspl{plc} are presented in this paper. 

In addition to the complexity of the algorithm to be offloaded, there are other parameters, such as network delay, that must be taken into account when calculating the break-even points. Using a previous work \cite{wfcs2021} and the research in this paper, each fraction of the break-even point calculation is identified.

Accordingly, the following contributions can be found in this paper:
\textbf{
\begin{itemize}
    \item Motivation for using computation offloading at \glspl{plc} to facilitate Industry 4.0 objectives.
    \item Proposal of a factory scenario, and identification of the \gls{rtt} of different communication systems.
    \item Derivation of break-even points for two different \glspl{plc}, based on different communication technologies and open communication interfaces and protocols. 
\end{itemize}
}

Therefore, the paper is structured as follows: Sec. \ref{sec:Related Work} gives an overview about related work on this topic, while limitations of \glspl{plc} and benefits that can be achieved by applying computation offloading are presented in  Sec. \ref{sec:Need for Computation Offloading}. Moreover, Sec. \ref{sec:Communication Interfaces} lists relevant communication interfaces that are suitable for computation offloading. In addition, the factory scenario, which serves as basis for our break-even point calculations, as well as \glspl{rtt} of different communication systems are proposed in Sec. \ref{sec:Factory Scenario}. Furthermore, Sec. \ref{sec:Evaluation} evaluates the break-even points that are also derived in this section. Finally, a conclusion is given (Sec. \ref{sec:Conclusion}).

\section{Related Work}%
\label{sec:Related Work}

For the realization of smart manufacturing, cloud computing is seen as one of the key technologies \cite{7571077}.  In particular, many mobile devices appear that require high-performance wireless communications that differ from today's applications. Therefore, an architecture that can meet these requirements was introduced \cite{gundall2021introduction}. Since most mobile devices have limited resources in terms of computation and energy, \cite{8254628} proposes a computation offloading approach to improve the energy consumption of mobile devices. This approach can be used to extend the operating time for battery-powered systems. However, resource-constrained devices, such as \glspl{plc}, are also found in the field level of industrial systems. Since these devices do not have any mobility, they are connected to all systems wireline. Therefore, battery life is not the focus when considering resource offloading. Here, especially the offloading of complex algorithms can bring advantages in terms of a lower processing time. Furthermore, offloading algorithms do not necessarily have to bring quantitative advantages, but can also enable advanced functions. For this reason, \cite{gundall2020introduction} proposed several functions that enable seamless reconfiguration and redeployment of virtualized process controllers, while maintaining the required availability in the industrial environment, which is quite different from other domains. Moreover, \cite{gundall2021feasability} demonstrates the feasibility of the proposed concept. Even if there are various reasons for applying computation offloading at \glspl{plc}, we focus on the performance, by identifying break-even points for the processing time. Here, the authors in \cite{wfcs2021} have already assessed open interfaces and protocols of \glspl{plc}. 
\section{Motivation for Computation Offloading}
\label{sec:Need for Computation Offloading}
The first part of this section describes the state of the art process control and the limitations given by the utilization of conventional \glspl{plc}, while Section \ref{subsec:Benefits by Computation Offloading} explains the benefits, given by the application of computation offloading.
\subsection{State of the Art Process Control}%
\label{subsec:State of the Art Process Control}

\begin{figure}[tb]
\centerline{\includegraphics[width=0.8\columnwidth, trim = 160 70 350 120,clip]{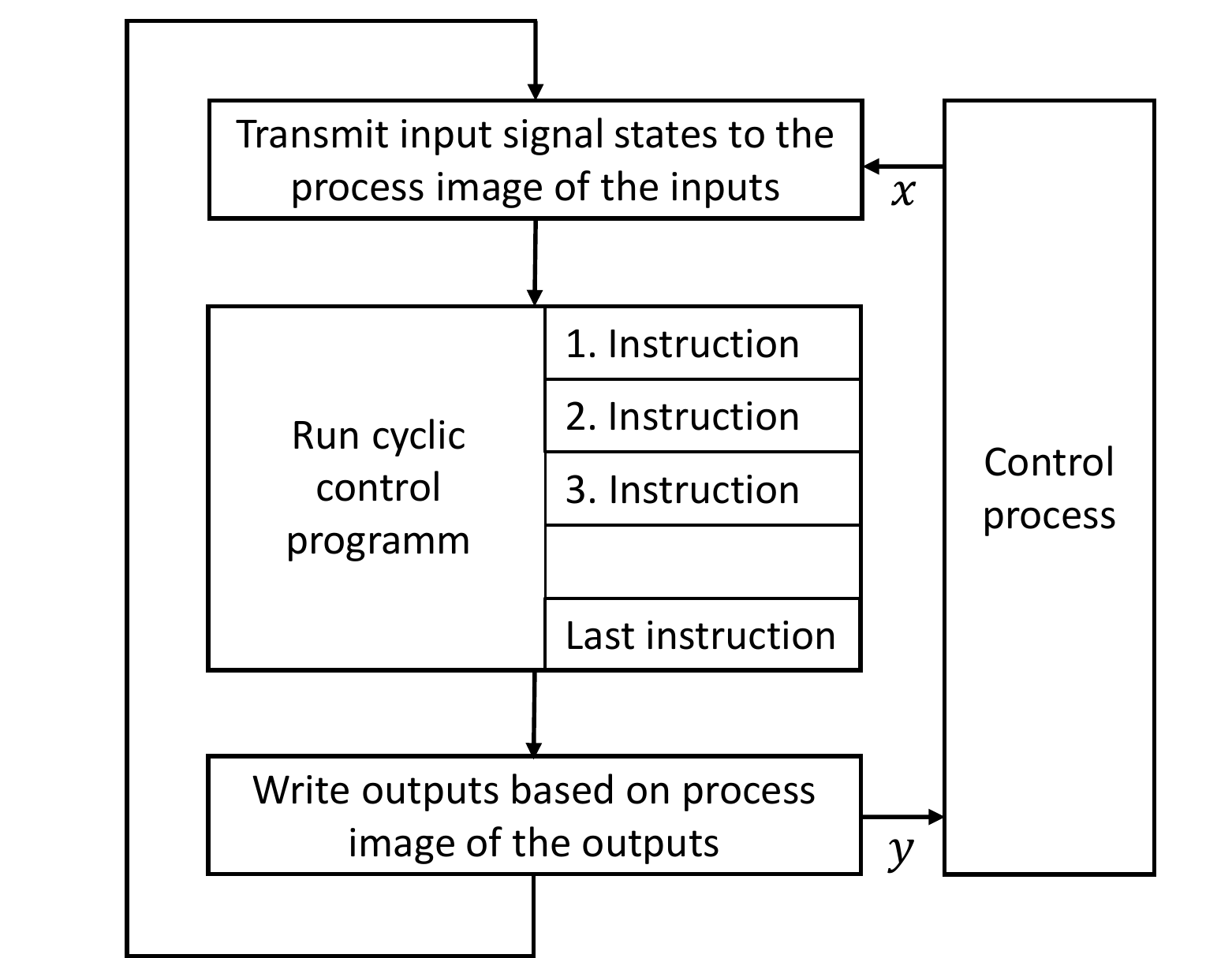}}
\caption{State and sequence illustration of a cyclic \gls{plc} program}
\label{fig:State and sequence illustration of a cyclic plc program}
\end{figure}

Today's industrial plants are controlled by \glspl{plc} \cite{6246692}. A \gls{plc} is a special microprocessor-based control unit that is known for a deterministic and \gls{rt} program execution and communication. Thus, these  controllers are highly sophisticated for continuous control tasks. Therefore, Fig. \ref{fig:State and sequence illustration of a cyclic plc program} shows the sequence for both program processing and data exchange.
Each cycle starts with the reading of the input values, which are transmitted in the so-called “process image of the inputs”. This means that the values are fixed for the rest of the iteration of the control program. This step is important for a deterministic behavior, because changes from the sensor values during the calculations do not influence the control program for the recent cycle. In the next step, the program logic defines the correspondent output values and writes them into the “process image of the outputs”, which is than transmitted to the output modules. Afterwards, the next iteration starts.

For addressing \gls{rt} aspects, the cycle time of the aforementioned process is a suitable \glspl{kpi}, where the cycle time of a \gls{plc} depends on the number of inputs, number of outputs, and complexity of the executed algorithms. Consequently, raising the complexity of algorithms, e.g. for an improved closed-loop control, increases the cycle time of both, the \gls{plc} and the closed-loop application. As the complexity of algorithms in Industry 4.0 applications increases, such as for image recognition and \gls{ml}, the computational power of \glspl{plc} is not sufficient. To simulate such a scenario, we built a load generator that calculates digits of the mathematical constant $\mathrm{\pi}$ using Leibniz formula, which is shown in Eq. \ref{eq:leibniz}: 
\begin{equation}
\label{eq:leibniz}
\pi = 4 \cdot \sum_{k=0}^n \frac{(-1)^k}{2k+1}
\end{equation}
Here, it is important to mention that $n$ does not represent the digits of $\mathrm{\pi}$, but the partial sums of the Leibniz series. Therefore, Tab. \ref{tab:2} gives shows the required values for $n$, in order to obtain the first six digits. 
\begin{table}[tb]
\caption{Required number of partial sums for the calculation of $\mathrm{\pi}$}
\begin{center}
\begin{tabulary}{\columnwidth}{|C|C|C|}
\hline
\textbf{$\pi$} & \textbf{\textit{n}} & \textbf{FLOP} \\
\hline
 3 & 2 & $\approx$10\textsuperscript{1}\\
\hline
 3.1 & 32 & $\approx$10\textsuperscript{3}\\
\hline
 3.14 & 1,000 & $\approx$10\textsuperscript{6} \\
\hline
 3.141 & 10,000 & $\approx$10\textsuperscript{8}\\
\hline
 3.141 & 100,000 & $\approx$10\textsuperscript{10} \\
\hline
 3.14159 & 1,000,000 & $\approx$10\textsuperscript{12}  \\
\hline
\end{tabulary}
\label{tab:2}
\end{center}
\end{table}

\begin{figure}[b!]
\centerline{\input{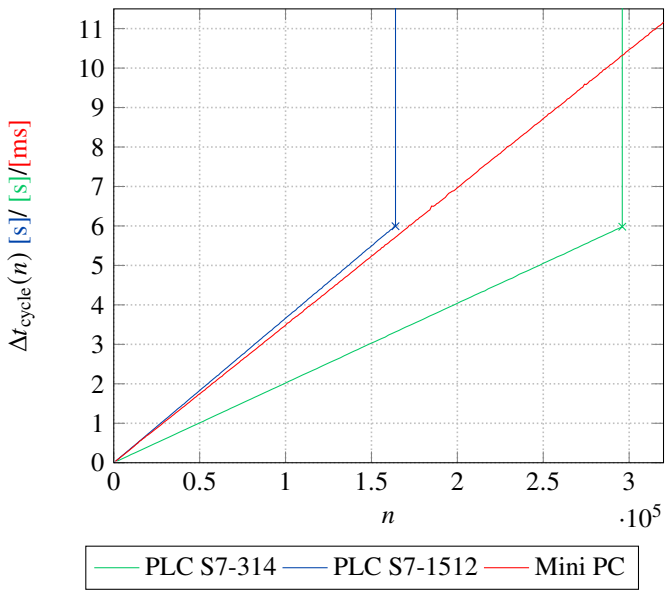}}
\caption{Dependency between the number of calculations and the cycle time of two typical S7 \glspl{plc} compared to a mini PC}
\label{fig:Dependency between the number of calculations and the cycle time of the PLC}
\end{figure}

Furthermore, the corresponding number of \glspl{flop}\footnote{We do not differentiate between addition, subtraction, multiplication, and division.} is listed, calculated with formula Eq. \ref{eq:flop} that determines the \glspl{flop} of the Leibniz series dependant on the upper bound of the sum $n$.
\begin{equation}
\mathrm{FLOP} = [n \cdot (n+\mathrm{3})]+1 = n^2+3n+1 \label{eq:flop}
\end{equation}
To investigate the influence of $n$ to the performance of the \gls{plc}, the load generator increases $n$ in each cycle. Fig. \ref{fig:Dependency between the number of calculations and the cycle time of the PLC} illustrates the dependency between $n$ and the difference cycle time $\Delta t_\mathrm{cycle}(n)$ for two representative \glspl{plc} that are listed in Tab. \ref{tab:equipment}. 

The plot shows that the dependency between these parameters are linear functions, until the point where the \gls{plc} goes in stop state. Thus, we have the following correspondence, where $\Delta t_\mathrm{cycle}(n)$ refers to the increased cycle time, $n$ to the number of calculated partial sums, and $\mathrm{c}$ is the constant time that is needed for each iteration of Eq. \ref{eq:leibniz}:

\begin{equation}
\label{eq:dependency between cycle and calculations}
\Delta t_\mathrm{cycle}(n) = \mathrm{c} \cdot n  
\Rightarrow \mathrm{c}=\frac{\Delta t_\mathrm{cycle}(n)}{n}
\end{equation}

\begin{table}[tb]
\caption{Hardware configurations}
\begin{center}
\begin{tabular*}{\columnwidth}{|c|p{0.34\columnwidth}|p{0.09\columnwidth}|p{0.18\columnwidth}|}
\cline{1-4}
\textbf{Equipment}  & \textbf{Specification} & \multicolumn{2}{|c|}{\textbf{Constant}} \\
  &  & \textbf{Name} &  \textbf{Value [ms]}\\
\cline{1-4}
PLC & SIMATIC S7 CPU  & c\textsubscript{1}  & $3.65 \cdot 10^{-2}$\\
S7-1512 & 1512SP F-1 PN for ET 200SP, Fw. V2.8 &  &\\
\cline{1-4}
PLC & SIMATIC S7 CPU  & c\textsubscript{2}  & $2.02 \cdot 10^{-2}$\\
S7-314 & 314C-2 PN/DP, Fw. V3.3  &  &\\
\cline{1-4}
mini PC &  Intel Core i7-8809G, 2x16 GB DDR4, Intel i210-AT Gibgabit NIC, Ubuntu 18.04 LTS 64-bit, \linebreak Linux 4.19.103-rt42 & c\textsubscript{3}  & $3.49 \cdot 10^{-5}$ \\
\cline{1-4}
\end{tabular*}
\label{tab:equipment}
\end{center}
\end{table}

Therefore, Fig. \ref{fig:Dependency between the number of calculations and the cycle time of the PLC} depicts the limits of conventional \glspl{plc}. In this case, the maximum value for the partial sums are $n_\mathrm{max} = 1.64\cdot 10^5$ for S7-1512 and $n_\mathrm{max} = 2.96\cdot 10^5$ for S7-314 \gls{plc}. This corresponds to the first five digits of $\pi$ for S7-1512, and six digits for S7-314, respectively. If $n_{max}$ is increased further, the \gls{plc} goes into a special stop state. This means, that the plant is going to be stopped. Therefore, this condition has to be avoided by adding more \gls{plc} \glspl{cpu} or transferring the task to an edge node in cases there is a high load on the \gls{cpu} of the \gls{plc}.

\subsection{Benefits by Computation Offloading}%
\label{subsec:Benefits by Computation Offloading}
Besides the computation performance of the selected \glspl{plc}, Fig. \ref{fig:Dependency between the number of calculations and the cycle time of the PLC} also shows the execution time of Eq. \ref{eq:leibniz} for the same values of $n$ of a \gls{cots} mini PC that is also listed in Tab. \ref{tab:equipment}. We decided to use a mini PC for the comparison because it is comparable to the used \glspl{plc} in terms of space consumption and pricing. Fig. \ref{eq:dependency between cycle and calculations} indicates that the cycle time of the mini PC is $\approx 1 \cdot 10^{-3}$ lower compared to both \gls{plc}. This means that the computation performance of the small-sized PC has already the same  computational performance as 1,000 \glspl{plc}. Thus, computation offloading can achieve a reduction in the cycle time of algorithms with a certain complexity. The criterion that must be satisfied so that there is a benefit is expressed in Eq. \ref{eq:overall}, where $t_\mathrm{ro}$ stands for the overall time consumed by resource offloading and $\mathrm{\Delta} t_\mathrm{cycle}$ for the corresponding increase of the cycle time of the \gls{plc}. 

\begin{equation}
t_\mathrm{ro}(n) < \Delta t_\mathrm{cycle}(n) \label{eq:overall}
\end{equation}

\begin{figure}[tb]
\centerline{\includegraphics[scale=1.0]{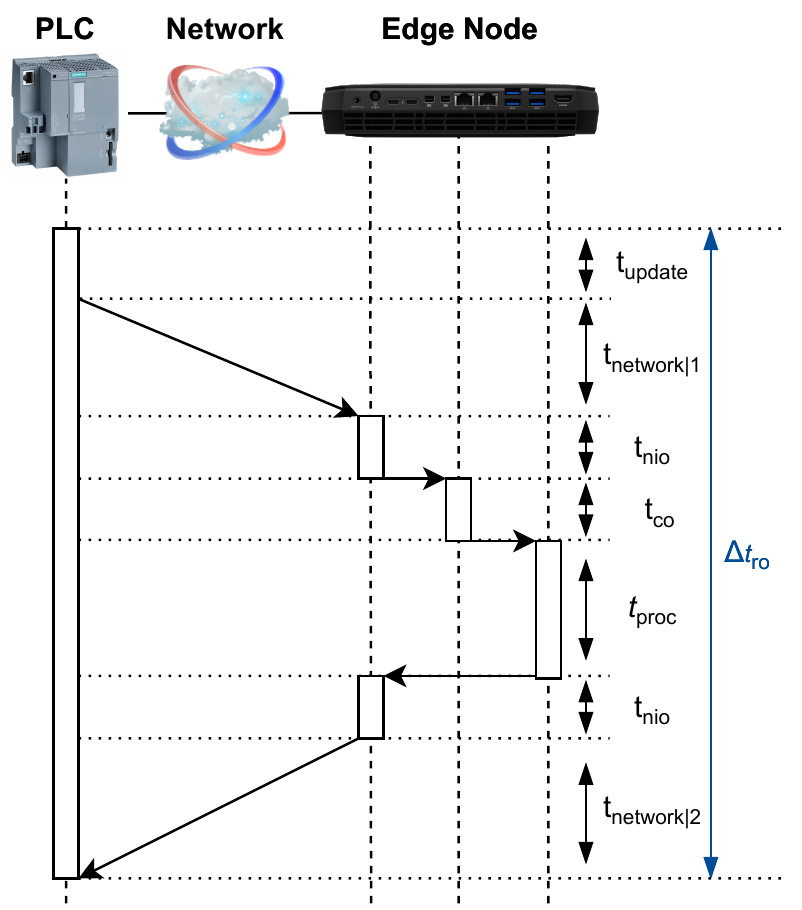}}
\caption{Sequence Diagram for the application of computation offloading.}
\label{fig:Sequence Diagram}
\end{figure}

Therefore, the time consumed by offloading calculations can be divided up further, as shown in Fig. \ref{fig:Sequence Diagram} and mathematically expressed in Eq. \ref{eq:ro}.

\begin{equation}
t_\mathrm{ro}(n) = \mathrm{t_{update}} + \mathrm{t_{network|1}} + 2 \cdot \mathrm{t_{nio}} + \mathrm{t_{co}} + t_\mathrm{proc}(n) +  \mathrm{t_{network|2}} 
\label{eq:ro}
\end{equation} 

It can be seen that a high latency of the communication network has a negative impact on a closed-loop application that requires input data packets with a low latency. In addition, the update time plays a major role as it indicates the frequency with which data packets can be sent. It is defined as the \textit{"[...] time interval between any two consecutive messages delivered to the application."} \cite{3gppts22104}. 
This value is characteristic for the investigated \gls{plc}, but is not network dependent.  Moreover, $t_\mathrm{ro}$ depends on the delay of the network used for offloading the data. This is expressed by the delay of the forward path $\mathrm{t_{network|1}}$ and the backward path $\mathrm{t_{network|2}}$. In order to simplify Eq. \ref{eq:ro}, a variable $\mathrm{t_{network}}$ that is representing the total network delay, is used. As in a realistic scenario the algorithm is not processed on bare-metal, there is also an overhead due to the use of virtualization technology for both the network interface ($\mathrm{t_{nio}}$) and computation ($\mathrm{t_{co}}$). The overall overhead caused by virtualization use can be calculated with the help of Eq. \ref{eq:overhead}.

\begin{equation}
\mathrm{t_\mathrm{overhead}} = 2 \cdot \mathrm{t_\mathrm{nio}} + \mathrm{t_\mathrm{co}}
\label{eq:overhead}
\end{equation} 

Finally, the time required to process the algorithm is one of the summands of $t_\mathrm{ro}$. Taking into account the simplifications for network delay and virtualization overhead, Eq. \ref{eq:formula_ro_simplified} is a simplified version of Eq. \ref{eq:ro}.

\begin{equation}
t_\mathrm{ro}(n) = t_\mathrm{proc}(n) + \mathrm{t_{network}} + \mathrm{t_{update}}  + \mathrm{t_{overhead}}  \label{eq:formula_ro_simplified}
\end{equation} 

\section{Open Communication Interfaces of PLCs}
\label{sec:Communication Interfaces}
In order to realize the offloading of algorithms, the communication interfaces and protocols of the devices under consideration must also be taken into account. As already mentioned, the update time is a suitable metric. Since Industry 4.0 requires data exchange between \gls{it} and \gls{ot} (\gls{it}-\gls{ot} convergence), only open interfaces that use standard Ethernet and \gls{ip} layer are considered. Due to the fact, that most relevant communication interfaces of two representative \glspl{plc} were already identified and assessed in \cite{wfcs2021}, we shortly list the findings that are relevant for this work. Hence, Tab. \ref{tab1} lists the investigated communication interfaces as well as the values for the update time for three different message sizes.
\begin{table*}[tb]
\caption{Comparison of the most common and open interfaces of two representative \glspl{plc} \cite{wfcs2021}.}
\begin{center}
\begin{tabular*}{.9\textwidth}{|c|c|c|c|c|c|c|c|c|}
\cline{1-9}
\multicolumn{2}{|c|}{\textbf{Interface}}  & \textbf{Protocol} &  \multicolumn{6}{|c|}{\textbf{Min. Update Time [ms]}} \\
\cline{4-9}
\multicolumn{2}{|c|}{\textbf{Configuration}}  &   & \multicolumn{2}{|c|}{\textbf{1 Data Value}} & \multicolumn{2}{|c|}{\textbf{10 Data Values}} & \multicolumn{2}{|c|}{\textbf{100 Data Values}}\\
\multicolumn{2}{|c|}{} &  &\textbf{\textit{S7-314}} &\textbf{\textit{S7-1512}}  & \textbf{\textit{S7-314}} &\textbf{\textit{S7-1512}}  &\textbf{\textit{S7-314}} &\textbf{\textit{S7-1512}} \\
\cline{1-9} 
\multicolumn{2}{|c|}{Open User} & UDP &  1.00 & 3.61 &  1.00 & 3.60  & 1.00 & 3.63 \\
\cline{3-9} 
\multicolumn{2}{|c|}{Communication} & TCP & 1.01 &3.77 &  1.04 & 3.78 & 1.02  &3.83 \\
\cline{1-9} 
\multicolumn{2}{|c|}{LIBNODAVE} & ISO on TCP & 2.00 &1.32 & 2.00 & 1.32 & 4.00$^{\mathrm{2}}$ & 1.40\\
\cline{1-9} 
OPC UA & Write Service & UATCP &  -&6.83 & -& 7.36 & - &16.56\\
\cline{2-9} 
Server Client & Read Service & UATCP  &-& 9.11 & -& 30.35 & -& 246.1\\
\cline{1-9}
\multicolumn{2}{|c|}{OPC UA PubSub} & UADP &  -&1.02 & - & 1.26 &  -& 2.30$^{\mathrm{1}}$\\
\cline{1-9} 
\multicolumn{9}{l}{$^{\mathrm{1}}$Due to a bug in the recent PLC's firmware, where only 20 data values can be send, this value was estimated.} \\
\multicolumn{9}{l}{$^{\mathrm{2}}$For the S7-314 series this value is doubled, because two requests have to be sent.} \\
\end{tabular*}
\label{tab1}
\end{center}
\end{table*}
Since these interfaces and the measurements are described in \cite{wfcs2021} in detail, we only summarize the results.

\subsection{\gls{ouc}}
The so-called \gls{ouc} allows to send and receive user defined data packets that can use several \gls{ip}-based protocols. Here, \gls{udp} (RFC 768) and \gls{tcp} (RFC 793) were identified as most suitable. Due to the fact, that the user data is directly packed into the data packet, no additional protocol header has to be generated. This makes this interface very efficient, resulting especially in a high performance of the S7-314 \gls{plc}. Compared to this \gls{plc}, the minimal update time of this interface using S7-1512 \gls{plc} is much higher. However, it can be seen, that the value does not increase a lot for larger data sets that are send using this interface for both \glspl{plc}.  



\subsection{LIBNODAVE}
LIBNODAVE is a free and open source library for using the ISO-on-\gls{tcp} protocol (RFC 1006) that is also known as “S7~Protocol" and uses port 102 for communication \cite{hergenhahn2011libnodave}. Furthermore, the edge device has to request the data from the \glspl{plc} by sending so-called \textit{Job} messages, that are responded by an \textit{Ack\_Data} message. Here, it should be noted, that the \gls{pdu} size of the S7-314 \gls{plc} is limited to 240 bytes. Therefore, only $\approx$ 50 data values can be read within one request. Hence, for sending 100 data values, two requests have to be send. Thus, the update time for 100 data values is doubled for this \gls{plc}. Furthermore, it can be seen, that LIBNODAVE is less efficient compared to \gls{ouc} for S7-314 \gls{plc}. This may be a result of the additional protocol overhead, given by the S7~Protocol. However, for S7-1512 \gls{plc}, the value is lower with respect to both, S7-314 \gls{plc} and \gls{ouc} interface. This makes it a suitable interface for applying computation offloading for S7-1512 \gls{plc}.

\subsection{\gls{opcua}}

In order to facilitate the convergence between \gls{it} and \gls{ot}, \gls{opcua} \cite{IEC625411} was introduced. It aims at a secure, simple and platform-independent exchange of information between industrial applications \cite{leitner2006opc}. Since these specifications are not supported by S7-300 series, no readings are available for this \gls{plc} type. However, the usage of gateways that are connected close to the \gls{plc} are a possible solution, if \gls{opcua} is required as communication interface between \gls{plc} and edge device \cite{6009198}. 

\subsubsection{OPC UA Server Client}
The \gls{opcua} server client pattern supports the binary \gls{tcp}-based communication protocol (UATCP), which uses port 4840, and is well suited for embedded devices, such as \glspl{plc}. If data should be send from the \gls{plc} to the edge node using \gls{opcua} server client, two different roles can be assumed. If the \gls{plc} is the client, it can send data to the edge node using so-called \textit{WriteService}. In contrast, if the \gls{plc} is the server, the edge node must query the data values using \textit{ReadService}. The latter can be compared to the LIBNODAVE interface. It can be seen, that using \gls{opcua} client server model, the minimum update times of S7-1512 \gls{plc} using either \textit{WriteService} or \textit{ReadService} are higher compared to both other interfaces. This is a result of the big protocol overhead that brings up a lot of meta information \cite{wfcs2021}. Therefore, this interface can be important for a lots of use cases, but is not recommended if only the performance is relevant. 

\subsubsection{\gls{opcua} \gls{pubsub}}
In addition to the server client model, the \gls{pubsub} pattern was introduced in part 14 of the \gls{opcua} specifications \cite{IEC625414}. It allows devices to subscribe for messages that are published by other devices. Here, mappings to different well-known broker-based protocols, such as \gls{mqtt} and \gls{amqp}, as well as a custom \acrshort{udp}-based distribution that is called UADP are defined. Latter, that is based on the \acrshort{ip} standard for multicasting, is recently the only one supported by the \gls{plc} and is used for the investigations. Even if the value for the update time of 100 data values had to be estimated, it is visible that the update time increases with higher data sets. However, for 1 and 10 data values it is the most efficient interface of S7-1512 \gls{plc} making it a good candidate for offloading data to an edge device.

\section{Factory Scenario}
\label{sec:Factory Scenario}
\begin{figure}[tb]
\centerline{\includegraphics[width=\columnwidth]{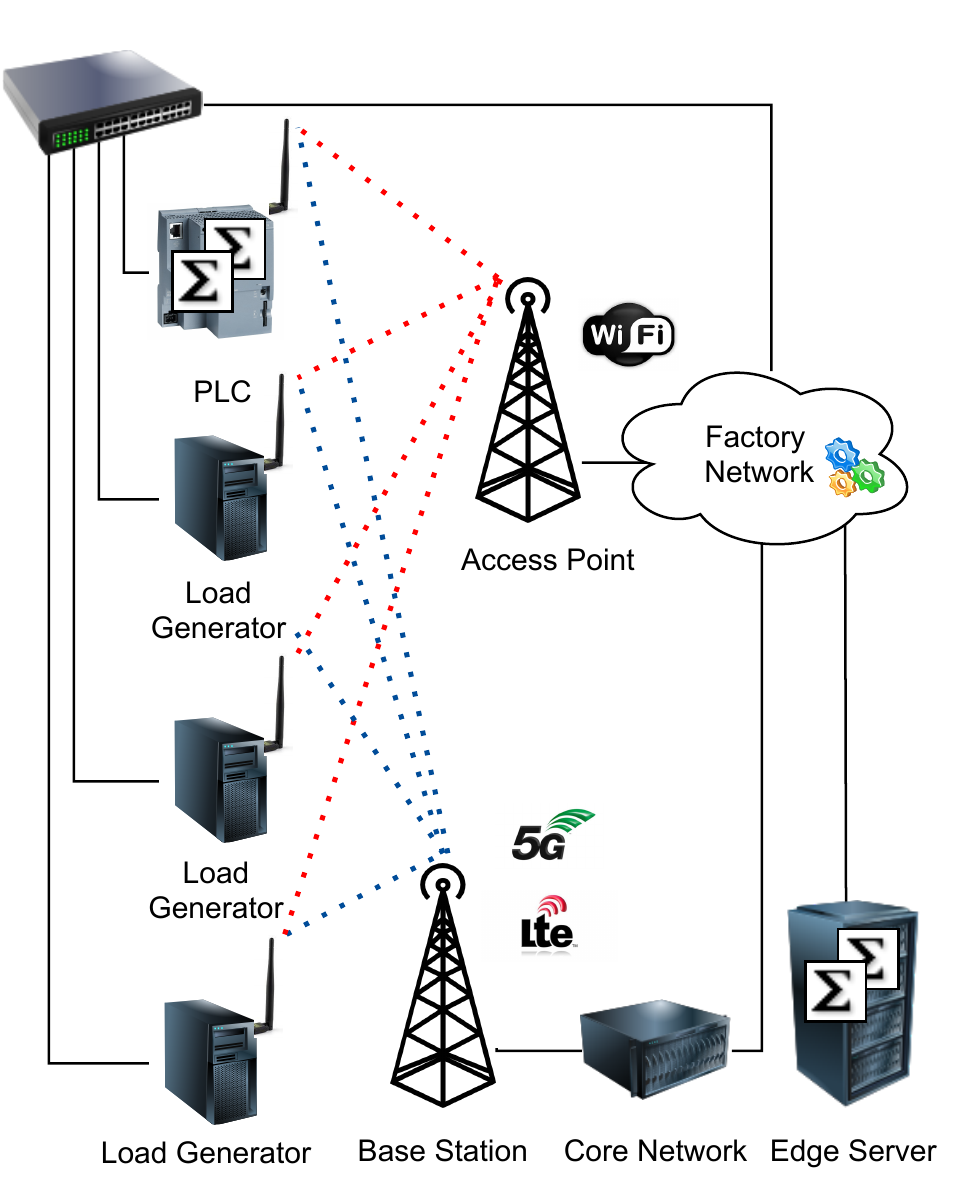}}
\caption{Factory scenario for the application of computation offloading, including both wireline and wireless data paths and network traffic generators.}
\label{fig:Factory scenario}
\end{figure}

In this section a factory scenario is proposed, on the basis of which the break-even points are derived in the following section (Sec. \ref{sec:Evaluation}). Since the calculated break-even points are only valid for this specific scenario, it should be as realistic as possible. Thus, the factory scenario, which serves as basis for our investigations, is shown in Fig. \ref{fig:Factory scenario}.
Since in a realistic factory scenario are other devices in the factory network besides the \gls{plc} and the edge node itself, we integrate three load generators that produce network traffic. Furthermore, we assume the following parameters for the factory network, which are also listed in  Tab. \ref{tab:constants}.

\begin{table}[tb]
\caption{Estimations for the factory network and constants for the break-even point calculations.}
\begin{center}
\begin{tabulary}{\columnwidth}{|C|C|C|}
\hline 
\textbf{Estimated constant} & Constant & \textbf{Value} \\
\hline
line length & l & 1,000 m \\
number of hops & z & 10 \\
network delay & $\mathrm{t_{network}}$ & 80 µs \\
virtualization overhead computation & $\mathrm{t_{co}}$ & 1 µs \\
virtualization overhead networking & $\mathrm{t_{nio}}$ & 1.5 µs \\
virtualization overhead & $\mathrm{t_{overhead}}$ & 4 µs \\
\hline
\end{tabulary}
\label{tab:constants}
\end{center}
\end{table}


Since the latency of a wired network depends mainly on the length of the line and the hops from the source to the receiver, but not necessarily on the complexity of the offloaded algorithm, it can be assumed to be constant for a given network. The authors in \cite{gundall2020application} proposed a factory scenario that forms the basis for our assumptions. They found that at least 1-3 hops are realistic to achieve an application running on an edge node. To make a conservative assumption, we include 2 more hops. Furthermore, we assume a cable length of 500 m. This means that a maximum of 500 m of cable and 5 switches or routers in the forward and reverse paths are assumed, giving a total line length of 1,000~m and the number of total hops of 10. Since the overhead of a standard 8-port network switch was identified as $\approx$15 µs in the investigations in \cite{gundall2020application}, we assume the one-way delay to be 7.5~µs. 
In addition, the investigations in \cite{gundall2020application} identified the overhead of container virtualization technology compared to bare-metal. They assume $<$1 µs for the computation overhead and $\approx$ 1.5~µs for the network interface overhead with proper setup and configuration. The total overhead given by virtualization is thus assumed to be 4~µs. 

Furthermore, for offloading data from a \gls{plc} to an edge node, which is mainly required for brownfield scenarios, we assume the following possible scenarios:
\begin{enumerate}
    \item A \gls{lan} is available and the \gls{plc} is integrated wireline.
    \item No \gls{lan} is available and the \gls{plc} is integrated wireless.
    \begin{itemize}
        \item Wi-Fi is used for the integration.
        \item 4G/5G is used for the integration.
    \end{itemize}
\end{enumerate}

\begin{figure}[htbp]
\resizebox{\columnwidth}{!}{%
\input{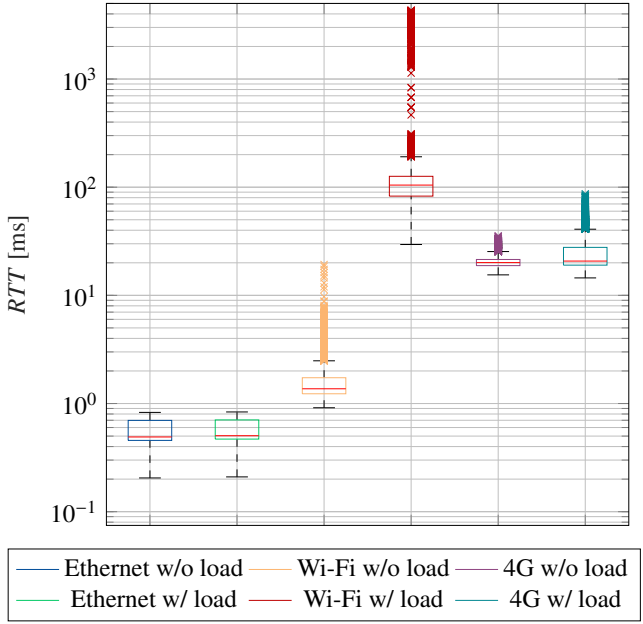}
}
	\caption{Readings of the \glspl{rtt} for the following three communication systems with and without additional load: (i) Ethernet, (ii) Wi-Fi, (iii) 4G.}
\label{fig:1}
\end{figure}

Since we assume all of the three possibilities as realistic, the \gls{rtt} of these communication systems should be identified and added to the network delay of the factory network. Therefore, we made 10,000 samples per measurement for a update time of 1~ms for each of the communication systems with and without additional load on the network. For the wireline setup we used a standard 8-port Ethernet switch, for the Wi-Fi connection a \gls{cots} IEEE~802.11ac Wi-Fi router, and for the 4G communication system a commercial 4G system that serves as \gls{npn}. The readings of the \glspl{rtt} for the specific communication system are showed in Fig. \ref{fig:1} and Tab. \ref{tab:2}.

\begin{table}[htbp]
\caption{Median and maximum values of the \glspl{rtt} for the investigated communication systems.}
\begin{center}
\begin{tabulary}{\columnwidth}{|p{0.13\columnwidth}|C|C|C|C|C|C|C|}
\hline
&\multicolumn{7}{c|}{\textbf{Round-trip time [ms]}} \\
\cline{2-8} 
 &\multicolumn{2}{c|}{\textbf{Ethernet}} & \multicolumn{2}{c|}{\textbf{\mbox{Wi-Fi}}} & \multicolumn{2}{c|}{\textbf{4G}} & \textbf{5G}\\
  & w/o & w/ & w/o & w/ & w/o & w/& -\\
  & load& load& load& load& load& load& - \\
   \hline
 \textbf{Median} &0.49& 0.50 & 1.37 & 104.60 & 20.08 & 20.75 & 2.00$^{\mathrm{1}}$\\
  \hline
 \textbf{Max} & 0.83 & 0.84 & 19.29 & 4320.88 & 35.58 & 86.95 & - \\
  \hline
  \multicolumn{8}{l}{$^{\mathrm{1}}$This value was adopted of the specification of the 3GPP.} \\
\end{tabulary}
\label{tab:2}
\end{center}
\end{table}


It can be seen, that using Ethernet as communication system, three traffic generators do not have a big impact on the \gls{rtt}, compared to an idle network. This means, that the \glspl{rtt} for the wireline setup are very low compared to the other communication systems, as expected. Looking at the Wi-Fi communication system, it can be seen, that the median value for the scenario without additional load on the network is also in the range of less than 1.4~ms. However, there are many outliers that are up to 15 times higher than the median value. If the values are compared to the load scenario, the values of the Wi-Fi systems are much worse. This is different in the 4G system. Here, the median values for the \glspl{rtt} are basically higher compared to both other systems, but the determinism is much higher than in the \mbox{Wi-Fi} system and the 100\% load does not have a big impact on the median value of the \gls{rtt}. Moreover, it can be seen that the maximum value of the \gls{rtt} for the 4G system is lower than the median value for the Wi-Fi system under load. Furthermore, since 5G communication systems are promising latencies of $<$1~ms for up- and downlink, we assumed a value of 2~ms as \gls{rtt} for a possible 5G deployment.

\section{Evaluation}
\label{sec:Evaluation}

After all the constants of Eq. \ref{eq:formula_ro_simplified} have been determined, the break-even points of the examined interfaces and scenario can be determined. For calculating the break-even points, the condition in Eq. \ref{eq:br_cond} has to hold. 

\begin{equation}
t_\mathrm{{ro}(n_{be})}  \stackrel{!}{=}  \Delta t_\mathrm{{cycle}(n_{be})} \label{eq:br_cond}
\end{equation}

Then, Eq. \ref{eq:dependency between cycle and calculations} can be inserted in Eq. \ref{eq:formula_ro_simplified} and resolved after $\mathrm{n_{be}}$, which leads to Eq. \ref{eq:br}.
\begin{equation}
\begin{split}
\mathrm{n_{be}} \cdot \mathrm{c_{plc}}  = \mathrm{n_{be}} \cdot \mathrm{c_{edge}} + \mathrm{t_{network}} + \mathrm{t_{update}}  + \mathrm{t_{be}}  \\\\
\Rightarrow \mathrm{n_{be}} = \frac{\mathrm{t_{network}} + \mathrm{t_{update}}  + \mathrm{t_{overhead}}}{\mathrm{c_{plc}}-\mathrm{c_{edge}}}
\end{split}
\label{eq:br}
\end{equation} 

\begin{figure}[htbp]
\centerline{\includegraphics[width=0.9\columnwidth, trim = 45 475 220 55,clip]{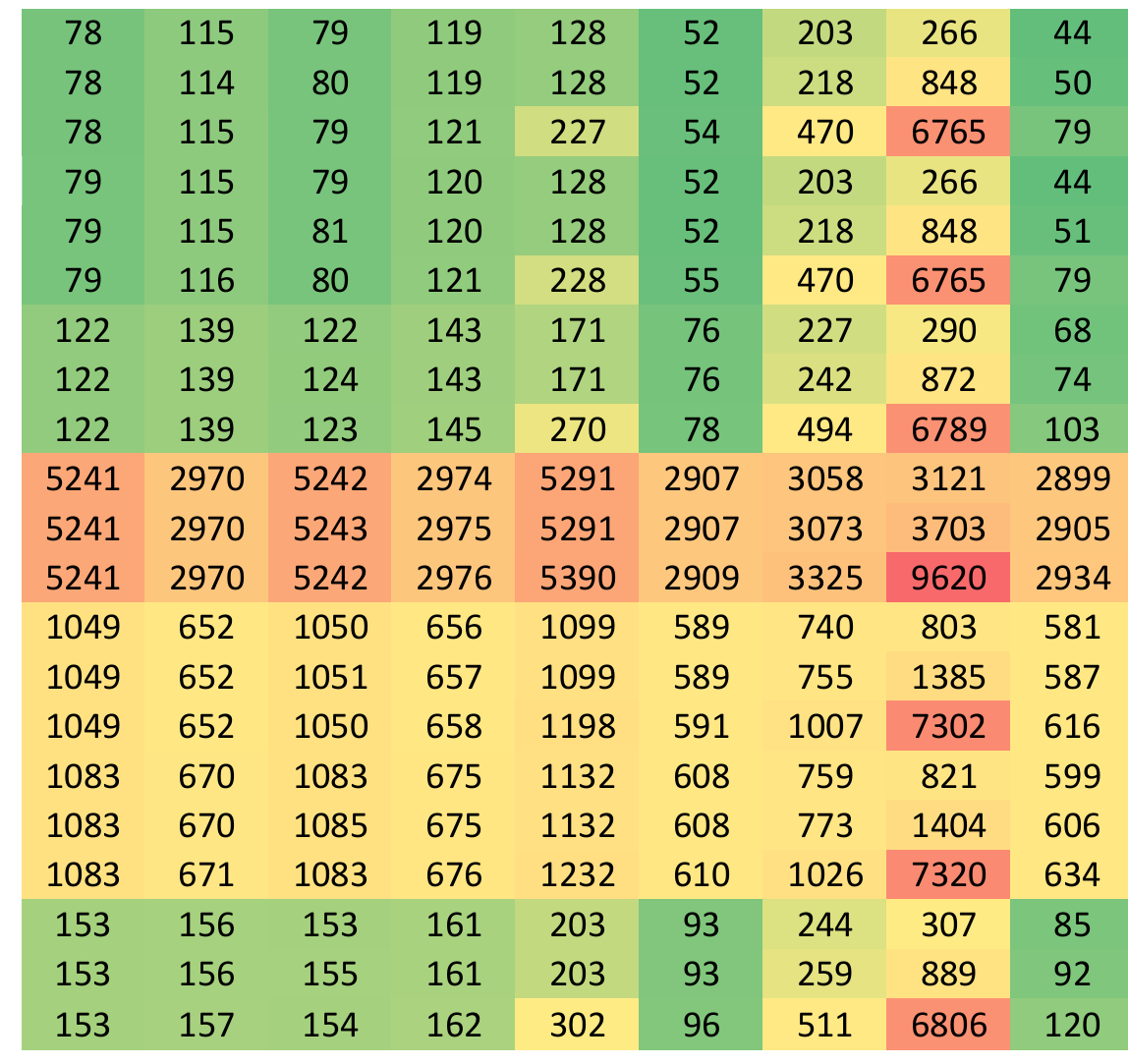}}
\caption{Heatmap for the calculated break-even points of Tab. \ref{tab: Breakeven}.}
\label{fig:heatmap}
\end{figure}

The results, which can be found in Tab. \ref{tab1}, indicate the complexity of a algorithm that is required to have a quantitative benefit for computation offloading, taking the specific network into account.  Additionally, a heatmap was produced out of the data and is show in Fig. \ref{fig:heatmap}
.
\begin{table*}[tb]
\caption{Break-even points for each of the investigated interfaces for the investigated \glspl{plc} and different communication systems.}
\begin{center}
\begin{tabular*}{\textwidth}{|c|c|c|c|c|c|c|c|c|c|c|c|c|}
\hline
\multicolumn{4}{|c|}{} & \multicolumn{9}{c|}{\textbf{Break-even points [$\mathrm{n_{be}}$]}} \\
\cline{1-13}
\multicolumn{4}{|c|}{\textbf{Interface $\rightarrow$}} & \multicolumn{4}{c|}{\textbf{Open User Communication}} & \multicolumn{2}{|c|}{\textbf{LIBNODAVE}} & \multicolumn{2}{c|}{\textbf{OPC UA Server Client}}&\textbf{OPC UA} \\
\multicolumn{4}{|c|}{\textbf{}} & \multicolumn{2}{c|}{\textbf{UDP}}&\multicolumn{2}{c|}{\textbf{TCP}} & \multicolumn{2}{|c|}{\textbf{}}& \textbf{Write Service} & \textbf{Read Service} & \textbf{PubSub}\\
\cline{1-13}
& & \textit{Load} & \textit{Data} & \textit{S7-} & \textit{S7-} & \textit{S7-} & \textit{S7-} & \textit{S7-} & \textit{S7-} & \textit{S7-} & \textit{S7-} & \textit{S7-}  \\
& &  $\downarrow$ & \textit{Values} $\downarrow$ & \textit{314} & \textit{1512}& \textit{314} & \textit{1512}& \textit{314} & \textit{1512}&  \textit{1512}&  \textit{1512}&  \textit{1512}\\
\cline{3-13}
\textbf{C} & \textbf{E} &  &\textit{1}                          &78&115&79&119&128&52&203&266&44 \\
\textbf{o} & \textbf{t} & \textit{w/o}& \textit{10}             &78&114&80&119&128&52&218&848&50 \\
\textbf{m} & \textbf{h} &&\textit{100}                          &78&115&79&121&227&54&470&6765&79 \\
\cline{4-13}
\textbf{m} & \textbf{e} &&\textit{1}                            &79&115&79&120&128&52&203&266&44 \\
\textbf{u} & \textbf{r}& \textit{w/} &\textit{10}               &79&115&81&120&128&52&218&848&51 \\
\textbf{n} & \textbf{n}&&\textit{100}                           &79&116&80&121&228&55&470&6765&79 \\
\cline{3-13}
\textbf{i} & \textbf{e}&&\textit{1}                             &122&139&122&143&171&76&227&290&68 \\
\textbf{c} & \textbf{t}&\textit{w/o} &\textit{10}               &122&139&124&143&171&76&242&872&74 \\
\textbf{a} & &&\textit{100}                                     &122&139&123&145&270&78&494&6789&103 \\
\cline{4-13}
\textbf{t} & \textbf{Wi}&&\textit{1}                            &5241&2970&5242&2974&5291&2907&3058&3121&2899 \\
\textbf{i} & \textbf{-}& \textit{w/} &\textit{10}               &5241&2970&5243&2975&5291&2907&3073&3703&2905 \\
\textbf{o} & \textbf{Fi}&&\textit{100}                          &5241&2970&5242&2976&5390&2909&3325&9620&2934 \\
\cline{3-13}
\textbf{n} & &&\textit{1}                                       &1049&652&1050&656&1099&589&740&803&581 \\
 & & \textit{w/o}&\textit{10}                                   &1049&652&1051&657&1099&589&755&1385&587 \\
\textbf{S} & &&\textit{100}                                     &1049&652&1050&658&1198&591&1007&7302&616 \\
\cline{4-13}
\textbf{y} & \textbf{4G}&&\textit{1}                            &1083&670&1083&675&1132&608&759&821&599 \\
\textbf{s} & & \textit{w/}&\textit{10}                          &1083&670&1085&675&1132&608&773&1404&606 \\
\textbf{t} & &&\textit{100}                                     &1083&671&1083&676&1232&610&1026&7320&634 \\
\cline{3-13}
\textbf{e }& &&\textit{1}                                       &153&156&153&161&203&93&244&307&85 \\
\textbf{m} & \textbf{5G}& - &\textit{10}                        &153&156&155&161&203&93&259&889&92 \\
\textbf{} & &&\textit{100}                                      &153&157&154&162&302&96&511&6806&120 \\
\cline{1-13}
\end{tabular*}
\label{tab: Breakeven}
\end{center}
\end{table*}

It is visible, that the application of computation offloading for S7-314 and S7-1512 \glspl{plc} is very efficient using \gls{ouc}. Thus, only the calculation of two digits of $\pi$ using Leibniz series on the edge device is beneficial, compared to processing it on the \gls{plc}. Furthermore, using LIBNODAVE is only slightly lower in performance compared to using \gls{ouc} for S7-314 \gls{plc} and much more efficient for S7-1512 \gls{plc}. Regarding \gls{opcua}, it can be seen, that server client model might beneficial in several aspects, which are explained in \cite{wfcs2021}, but \gls{pubsub} is superior regarding performance. 

Regarding the Wi-Fi connection, it is obvious, that for only more complex algorithms an advantage in terms of a lower overall processing time can be reached. Here, also the highest value is located, which requires an algorithm complexity of calculating approximately 4 digits of $\pi$ using Leibniz formula. This is related to $\approx$10\textsuperscript{8} \glspl{flop}. Besides this fact, it is visible, that the break-even points of the S7-1512 \gls{plc} for \gls{ouc} are now smaller compared to the S7-314 \gls{plc}. This effect is caused by the higher \glspl{rtt} of the communication system and the smaller computational power of the S7-1512 \gls{plc} and is also visible for the 4G system. 

Looking deeper into the values for the break-even points of the 4G system, the points where it is beneficial to offload an algorithm is approximately doubled for the S7-314 \gls{plc}, compared to S7-1512 for both \gls{ouc} and LIBNODAVE interface. 

Last but not least, the effect of the smaller computational power of S7-1512 \gls{plc} and the better performance regarding update time of S7-314 \gls{plc} are approximately compensated for \gls{ouc} using 5G as communication system.



\section{Conclusion}%
\label{sec:Conclusion}
In this paper, we proposed the motivation for the application of computation offloading at the field level of industrial plants using two representative \glspl{plc} as examples. Therefore, we elaborated on both their functionality and their limitations in terms of computational resources, interoperability, and flexibility compared to the computational resources of a potential edge node and the possibilities offered by virtualization technology. In addition, we listed relevant communication interfaces to access data of the \glspl{plc} and proposed a realistic factory scenario. Here, we assumed three different communication systems for the integration of a \gls{plc} in an \gls{it} environment. Moreover, we made measurements for these communication systems and the specific factory scenario and calculated the corresponding break-even points.




\balance
\printbibliography%
\nl%

%
%
\end{document}